\documentclass [a4paper,12pt]{article}
\usepackage[english]{babel}
\usepackage{graphicx,amsmath,amsfonts,amssymb,slashed,upgreek,color,caption,amscd,cite,cancel}
\usepackage{mathtools}

\newcommand{\be}{\begin{equation}}
\newcommand{\ee}{\end{equation}}
\newcommand{\bea}{\begin{eqnarray}}
\newcommand{\eea}{\end{eqnarray}}

\usepackage[stable]{footmisc}

\newcommand{\dee}{\hbox{\rm{d}}}
\newcommand{\pa}{\partial}
\newcommand{\lb}{\left[}
\newcommand{\rb}{\right]}
\newcommand{\lp}{\left(}
\newcommand{\rp}{\right)}
\newcommand{\la}{\left\{}
\newcommand{\ra}{\right\}}
\newcommand{\dpp}{\vcentcolon}
\newcommand{\bb}{\begin{eqnarray}}
\newcommand{\eee}{\end{eqnarray}}
\newcommand{\qq}{\quad}

\begin{document}

\begin{center}
\Large{\textbf{Minimal cosmography}} \\[0.5cm]
 
\large{Federico Piazza
 and Thomas Sch\"ucker}
\\[0.5cm]

\small{
\textit{CPT, Aix-Marseille University, Universit\'e de Toulon, CNRS UMR 7332\\ 13288 Marseille, France}}

\end{center}
\abstract{The minimal requirement for cosmography---a non-dynamical description of the uni\-verse---is a prescription for calculating null geodesics, and time-like geodesics as a function of their proper time. In this paper, we consider the most general linear connection compatible with homogeneity and isotropy, but not necessarily with a metric. A light-cone structure is assigned by choosing a set of geodesics representing light rays. This defines a ``scale factor" and a local notion of distance, as that travelled by light in a given proper time interval. We find that the velocities and relativistic energies of free-falling bodies decrease in time as a consequence of cosmic expansion, but at a rate that can be different than that dictated by the usual metric framework. By extrapolating this behavior to photons' redshift, we find that the latter is in principle independent of the ``scale factor". Interestingly, redshift-distance relations and other standard geometric observables are modified in this extended framework, in a way that could be experimentally tested. An extremely tight constraint on the model, however,  is represented by the blackbody-ness of the Cosmic Microwave Background. 
Finally, as a check, we also consider the effects of a non-metric connection in a different set-up, namely, that of a static, spherically symmetric spacetime.

\section{Introduction}

General relativity (GR) is the unique Lorentz-invariant low-energy theory of a massless spin-two field $g_{\mu \nu}$~\cite{Weinberg:1964ew}.  
The original semi-classical interpretation of $g_{\mu \nu}$ as the metric of spacetime emerges very convincingly in cosmology: the Hubble scale is where the corrections from a flat Minkowski metric become of order one. Basic cosmological tests and observations are probes of the geometry of the Universe. 

In this regard, it is worth reminding ourselves that all geometrical probes are ultimately ``time-like and null-like probes", whereas space-like distances are never directly  measured. 
Effectively, in static situations like the solar system, we can trade ``time-like" measurements for \emph{instantaneous} ones.  For example, the process of laying down a ruler many times to measure the distance between New York and Toronto, although described by a time-like curve, produces a well-defined \emph{space-like} distance as an outcome, because nothing else evolves meanwhile.

On the opposite, in a cosmological setup, because of the strong time dependence of the curvature invariants---in particular, those related with the Ricci-curvature---space-like distances are truly redundant.  They are part of the theory because calculable in terms of the metric elements, but they are operatively meaningless, unobservable.
For this reason,  it would be interesting to find some ingredient other than the metric that 
could \emph{weaken} the geometrical picture of GR, by encoding only the mutual time-like distances between the events---the latter are always directly measurable as proper time intervals of real observers. Such a weaker version of GR might show interesting departures from the actual theory precisely in cosmology, where time evolution is important.

While awaiting a breakthrough in this direction, here we content ourselves with weakening the metricity hypothesis down to the next obvious level, that of a general linear connection. One of the immediate drawbacks of this approach is the absence of a field theoretic description. For this reason, we do not attempt to write down equations of motions for such a connection. Rather, we impose a cosmological principle (homogeneity and isotropy) on the Christoffel symbols, and derive the kinematics of point particles following time-like geodesics.  Although not enough for cosmology, we show that these minimal ingredients are enough to set up a consistent and interesting \emph{cosmography}.

\section{Geometry}

We want to enforce homogeneity and isotropy on a linear connection $\Gamma^\lambda_{\mu \nu}$. A tensor field is symmetric under an infinitesimal transformation described by a killing vector $\xi^\mu$ if its Lie derivative in the $\xi$ direction is zero. As well known, Christoffel symbols are not tensors, but the covariant derivative of a vector field $A^\mu$ built with the Christoffel symbols, $\nabla_\mu A^\nu = \partial_\mu A^\nu + \Gamma^\nu_{\mu \alpha} A^\alpha$, is. Therefore we require that
\begin{equation}
L_\xi \nabla_\mu A^\nu\ = \ \xi^\alpha \partial_\alpha \nabla_\mu A^\nu + (\partial_\mu \xi^\alpha) \nabla_\alpha A^\nu - (\partial_\alpha \xi^\nu) \nabla_\mu A^\alpha \ = \ 0\, .
\end{equation} 
By requiring that $A^\mu$ itself has null Lie derivative with respect to $\xi$, we finally obtain a condition on the connection only,  
\begin{equation} \label{killing}
\xi^\alpha\partial_\alpha \Gamma^\lambda_{\mu \nu} - (\partial_\alpha \xi^\lambda) \Gamma^\alpha_{\mu \nu} + (\partial_\mu \xi^\alpha) \Gamma^\lambda_{\alpha \nu} + (\partial_\nu \xi^\alpha) \Gamma^\lambda_{\mu \alpha} + \partial_\mu \partial_\nu \xi^\lambda = 0\, .
\end{equation}

 Isotropy and homogeneity are expressed by a set of six  vector fields $J_i$ and $P_i$ (latin indexes $i, j, k \dots = 1,2,3$), with commutation relations
\begin{equation} \label{killing???}
[J_i, J_j] = \epsilon_{ijl} J_l, \quad [J_i, P_j] = - \epsilon_{ijl} P_l, \quad [P_i,P_j] = - k \epsilon_{ijl} J_l\, ,
\end{equation}
where $k = -1,0,1$ in the case where the group of symmetry is the hyperbolic, the Euclidean or the spherical one respectively. 
Next we want to choose a coordinate representation for such fields. A pretty standard one is
\begin{equation}
J_i = \epsilon_{i j l} x^j \partial_l, \qquad P_i = \sqrt{1- k r^2} \, \partial_i
\end{equation}
which, implicitly, also fixes the ``spatial" coordinates. In the above we have defined $r^2 = x_1^2 + x_2^2 + x_3^2$. It is easy to check that the above defined vectors satisfy the relations~\eqref{killing???}.

By applying the Killing equation~\eqref{killing} for the three generalized translations $P_i$, we obtain, after some algebra, the following equations
\begin{equation} \label{connection-eq}
(1 - k r^2) \partial_i \Gamma^\lambda_{\mu \nu} + k \delta_i^\lambda x_l \Gamma^l_{\mu \nu} - k (\bar x_\mu \Gamma_{i \nu}^\lambda + \bar x_\nu \Gamma_{\mu i}^\lambda) - \delta_i^\lambda\left(k \bar \delta_{\mu \nu} + k^2 \frac{\bar x_\mu \bar x_\nu}{1 - k r^2}\right) = 0\, ,
\end{equation}
where we used the bar to indicate quantities that are non zero only when their indexes are latin. In order to solve the above equation we will further demand the connection to be \emph{without torsion}, $\Gamma^\lambda_{\mu \nu}= \Gamma^\lambda_{\nu \mu}$, simply because, at the level of the kinematic description that we are after, we have not found anything interesting associated with it. 
By careful inspection, it is then possible to see that the most general torsionless solution of~\eqref{connection-eq} is
\begin{align}
& \Gamma_{00}^0 = q_2(t); \qquad \Gamma^0_{ij} = q_1(t) h_{ij}; \qquad \Gamma_{0j}^i = \Gamma^i_{j0} = q_3(t) \delta_j^i; \\ 
& \Gamma^0_{0i} = \Gamma^0_{i0} = \Gamma_{00}^i = 0; \qquad \Gamma_{jl}^i = k x^i h_{jl} \, ,
\end{align}
where $h_{ij}$ is the metric of the maximally symmetric 3-dimensional surface:
\begin{equation}
h_{ij} = \delta_{ij} + \frac{k\, x_i x_j}{1 - k r^2}\, .
\end{equation}
These results coincide with those found by  Goenner \& M\"uller-Hoissen \cite{goenner} (see also \cite{tomas} for related results).

The high degree of symmetry that we are considering fixes the connection in the spatial directions to be that of a maximally symmetric three-dimensional space, with a natural three-dimensional metric associated with it. The three dimensional line element of such space in spherical coordinates reads
\begin{equation}
ds^2 = \frac{dr^2}{1 - k r^2} + r^2 d\Omega\, .
\end{equation}
While $r$ characterizes the area of the two-spheres of symmetry, it is useful to introduce also an alternative radial coordinate that is ``flat" along the radial direction. In this case the metric takes the form
\begin{equation}
ds^2 = dR^2 + r^2(R) d\Omega\, ,
\end{equation}
where the relation between the two radii is
\begin{equation} \label{r(R)}
r(R) = \begin{cases} \frac{\sinh \left( \sqrt{|k|} R\right)}{\sqrt{|k|} }  \qquad \qquad \qquad \qquad & k<0 \\[2mm] 
R & k= 0\\[2mm]
\frac{\sin \left( \sqrt{|k|} R\right)}{\sqrt{|k|} } \qquad \qquad & k>0\, ,
\end{cases}
\end{equation}
and note that we are now allowing $k$ to take all values from $-\infty$ to $+\infty$. In this way, $k$ can be taken as a direct measure of the spatial curvature today. 

\section{Kinematics}

The equations for a geodesic $x^\mu(\tau)$ read
\begin{align}
\frac{d^2 x^\mu}{d \tau^2} + \Gamma^\mu_{\rho \sigma} \frac{d x^\rho}{d \tau} \frac{d x^\sigma}{d \tau} &= 0\, .
\end{align}
By requiring that the time coordinate $t$ be simply the proper time of the comoving observers, we can ask that for the particular geodesic for which the $x^i(\tau)$ are constant,  $t(\tau) = \tau$. This fixes $q_2 =0$. As for more general geodesics, it is not restrictive to concentrate on radial ones, thanks to the high degree of symmetry of the geometry that we are considering, $t(\tau), R(\tau)$. These two functions have to satisfy the equations
\begin{align}
\frac{d^2 t}{d \tau^2} &\, + \, q_1(t) \left(\frac{d R}{d \tau}\right)^2 \ = \ 0\, , \label{timeeq}\\ 
\frac{d^2 R}{d \tau^2} &\, + \, 2 q_3(t) \frac{d R}{d \tau} \frac{d t}{d \tau} \ =\ 0\, .
\end{align}
By denoting with a dot differentiation with respect to the time $t$, from a combination of the above equations we obtain
\begin{equation}\label{shift}
\frac{\ddot R}{\dot R} \ =  \ q_1 \dot R^2 - 2 q_3\, .
\end{equation}

Note that we are not equipped with a four dimensional metric and thus there is no such notion as a light cone so far. However, we can simply pick out a set of geodesics and baptize them ``light rays".  Let us  define them as 
\begin{equation} \label{secondorder}
\frac{d R}{d t} \ = \ \frac{1}{a(t)} \qquad \qquad {\rm (light\ rays)}\, .
\end{equation}
By requiring that these are geodesics, we obtain the relation\footnote{Note that in the usual (metric) case, $q_1 = a^2(t) H$, $q_3 = H$, where $a(t)$ is the scale factor and 
\\\indent\qq $H=\dot a/a$ the Hubble parameter.} 
\begin{equation}
q_1 = a^2 \left(2 q_3 - \frac{\dot a}{a}\right)\, .
\end{equation}
By setting the speed of light to one, we can then define the velocity of \emph{any} curve with respect to the cosmological reference frame as 
\begin{equation} \label{speed}
\boxed{v = a(t) \dot R\, .}
\end{equation}

We should emphasize that, in the present framework, $a(t)$ does not fully represent the scale factor as in a metric theory. This will be clear in what follows. Since it is related with the speed of light, if we are to define infinitesimal space distances through the time a light ray takes to travel them, we can say that the infinitesimal proper space distance is given by $a(t) dR$.

To highlight the reason why $a(t)$ is not ``completely" a scale factor, let us write the dynamical equation for the velocity of a test particle. By combining~\eqref{speed} and \eqref{shift} we obtain
\begin{equation} \label{slowdown}
\boxed{\frac{\dot v}{v} \ = \ - \, \frac{\dot b(t)}{b(t)}\, (1 - v^2)\, ,}
\end{equation}
where we have defined 
\begin{equation}
\frac{\, \dot b\, }{b} \ = \ 2 q_3 - \frac{\, \dot a\, }{a}\, .
\end{equation}
Note that, by the Lorentz transformation between the two times $t$ and $\tau$, $d \tau/d t = \sqrt{1-v^2}$, which is consistent with~\eqref{timeeq} upon use of 
~\eqref{slowdown}.

In the usual metric framework, $b=a$. Instead, here we have two, possibly different, ``scale factors". The first one, $a(t)$, converts comoving infinitesimal distances $dR$ into proper distances $d\sigma = a(t) dR$. Since comoving observers have fixed comoving coordinates, $a(t)$ tells the rate at which their mutual distances increase with time. But observers could as well be moving, say, with an initial separation $\Delta \sigma$ and with the same initial velocity both along the coordinate $R$. Because~\eqref{shift} is shift-invariant under translations in the $R$ coordinate, their mutual separation still grows with a factor of $a(t)$. This means that if the number of particles is conserved by the microphysics, their number density scales always as $a^{-3}$.

The other scale factor, $b(t)$, tells us the slow-down rate of the proper velocity of an object due to the expansion of the Universe. Crucially, $b(t)$, and not $a(t)$, is responsible for the loss of energy $E$ of a particle due to the expansion. For a particle of constant rest mass $m$, $E= m/\sqrt{1-v^2}$ and thus
\begin{equation} 
\frac{d \log E}{dt} \ = \ - \, \frac{\, \dot b\, }{b} v^2\, .
\end{equation}
For zero mass particles traveling at the speed of light, the above equation turns into an expression for the cosmological redshift,
\begin{equation}
\boxed{1+ z \ = \ \frac{b(t_0)}{b(t)}\, .}
\end{equation}
It might be useful to introduce an analogous quantity for $a$, 
\begin{equation}
1+ y \ = \ \frac{a(t_0)}{a(t)}\, .
\end{equation}
Therefore, the flux of conserved photons scales according to the  redshift $y$, whereas their individual energies redshift as $z$. 

We should emphasize that the above approach is legitimate 
if we decide to treat photons as particles and then take the limit where their velocity tends to the speed of light. Let us call this approach \emph{working hypothesis A} for definiteness. The alternative is to treat photons as waves of given frequency/wavelength and define their energy through the de Broglie's relation (\emph{working hypothesis B}). With this choice, the energy of the photons would redshift as $y= \frac{a_0}{a}-1$, and the latter would be the physical redshift. Not very surprisingly perhaps, we find it impossible to reconcile the wave-particle duality within this framework, simply because energies and lengths scale differently. Only for a metric connection do both working hypotheses \emph{A} and \emph{B} coincide. 

In the rest of the paper we will assume \emph{working hypothesis A}. The main reason is that doing so allows us to really consider an alternative, more general, cosmography. On the opposite, according to \emph{working hypothesis B} the scale factor $a$ does most of the work and there appears not to be any relevant departures from standard cosmography. Of course, the anomalous scaling of the energy for massive particles would imply an anomalous scaling of the peculiar velocities. \emph{Per se}, this phenomenon would be interesting to consider, but it would lead to much more subtle deviations from the standard picture, than applying the anomalous scaling of the energy all the way to the photons.

\section{Distances and Cosmological probes} \label{4}

Let's put ourselves at the center of the coordinate system $R=r=0$ and at the present time $t_0$. 
The comoving radius $R$ of a source that emitted a light ray at time $t$ and at redshift $y$, and reaches us now is
\begin{equation} \label{R}
R(y) \ = \ - \int_{t_0}^t \frac{dt'}{a(t')} \ = \  \int_0^y \frac{1}{H_a} dy\, ,
\end{equation}
where we have defined $H_a \equiv \dot a/a$.
The corresponding radial variable $r(y)$ is then obtained through~\eqref{r(R)}. 

The \emph{angular distance} of an object of real transverse size $\Delta L$, and that appears to cover an angle $\Delta \theta$ is the sky is defined as $D_A = \Delta L/\Delta \theta$. 
It is not difficult to convince oneself that 
\begin{equation}
D_A\ = \ \frac{r(y)}{1+y}
\end{equation}
Note that this quantity is formally identical to the usual expression upon substitution $y\rightarrow z$. Crucially, however, in a non-metric framework with our first definition for the photon energy, $y$ and $z$  are independent quantities and thus the relation between angular distance and redshift $D_A(z)$ will also be different. 

The \emph{luminosity distance} $D_L$ is defined through the total flux of energy from the emitting source. It is known  that three factors contribute in the relation between $D_L$ and $D_A$: the redshift of the energy of the single photon and the dilution of the number of photons per unit time and a geometric factor, see equation (3.31) of \cite{fleury}. With the first definition for the photon energy, these correspond to two different scaling laws. We conclude that the most general relation between $D_L$ and $D_A$ is indeed
\begin{equation} \label{ether}
\boxed{D_L\ = \ (1+z)^{1/2} (1+y)^{3/2}\, D_A\, .}
\end{equation}
From the above relation, we can also derive the general law for the total surface (\emph{Tolman test}) $B$, 
\begin{equation} \label{tolman}
B  \ \propto \ \frac{1}{(1+z)\, (1+y)^3}\, .
\end{equation}

For other geometrical tests such as the Alcock-Paczynski one~\cite{alcock}, one important quantity is the ``\emph{Hubble distance}" $D_H$, relating the proper length $\Delta L$ of an extended object along the line of sight with the difference in redshift $\Delta z$ of the two extremities of this object, $\Delta L = D_H \Delta z$. By taking the differential of $R$ defined in~\eqref{R} with respect to $z$ we find
\begin{equation}
D_H\ = \ \frac{1}{H_a} \frac{d y}{dz}\, .
\end{equation}

Another potentially interesting quantity is the \emph{redshift drift}, measured by looking at the redshift of a given source at two different times: $t_0$ and $t_0 + dt_0$.  The variation of the redshift is given by
\begin{equation}
\frac{d z}{d t_0} \ = \ H_b(t_0) (1+z) \, - \, H_b(z) \frac{1+z}{1+y}\, , 
\end{equation}
where we have defined $H_b \equiv \dot b/b$.

\section{Discussion}

The degree of non-metricity of our cosmology is measured by the difference between the two redshifts $y$ and $z$ (see also~\cite{Bassett:2013qqa}). The first is a measure of how much the photons dilute in a given volume because of the expansion. The second accounts for their individual energy loss. In a metric theory, these two aspects are inextricable, as expressed by the Tolman test~\cite{tol} and by the  Etherington  distance duality relation \cite{ethe}. Our equations~\eqref{ether} and~\eqref{tolman} are the appropriate generalizations of these formulae in the presence of a connection that  does not derive from a metric. More generally, the deviations from standard cosmology of the geometrical quantities calculated in Sec~\ref{4} are all given in terms of the difference between $y$ and $z$ and are, therefore, inter-related. It would be interesting to keep this  in mind when testing the consistency of the cosmological standard model with different probes.

Beyond the point-particle limit, our set-up lacks of a field-theoretic description. Nonmetricity
allows different scalings for positions and energies, somewhat mining the very
core of particle-wave duality.  It is thus not surprising that field theory rebels against this. Even more problematic is confronting this anomalous behavior with the spectrum of the photons  of the Cosmic Microwave Background (CMB), whose deviations from a perfectly blackbody have been constrained by COBE to be extremely tiny~\cite{Mather:1993ij,Fixsen:1996nj}.   It is not difficult to see that the form of the Planckian spectrum is preserved in time---and just temperature-rescaled---only if the number density of photons and their frequencies scale with the same scale factor $a$, as $a^{-3}$ and $a^{-1}$ respectively.  With this reasoning, violations of the distance-duality relation between last scattering and today were constrained to the level of $10^{-4}$ by the analysis of~\cite{Ellis:2013cu}. Analogous limits on deviations from metricity can be derived by solar system tests, as we show in the Appendix by looking at spherically symmetric and stationary connections. 

Despite these difficulties, we find the general possibility of anomalous cosmological scalings intriguing. 
The fact that we have to invoke cosmic acceleration twice (inflation and dark energy) in order to explain cosmological data seems, at times, to insinuate that we are still missing some major unifying ingredient in the description of the universe as a whole. On the other hand, the difficulties encountered by any alternative route show how solid and ``forced" our current theoretical understanding (here, in particular, metricity) is. Or, perhaps,  how much more creative we need to be. 

\section*{Acknowledgements}
 We would like to thank the referee for this constructive report, that allowed us to correct equation (\ref{ether}).
 
This work has been carried out in the framework of the Labex Archim\`ede (ANR-11-LABX-0033) and acknowledges the financial support of the A*MIDEX project (ANR-11-IDEX-0001-02), funded by the ``Investissements d'Avenir" French Government programme managed by the French National Research Agency (ANR).

\section*{Appendix}
 
So far we have generalised the Robertson-Walker metric to the setting where the gravitational field is encoded in a torsionless connection. In absence of a metric, but in presence of matter, there is no natural generalisation of the Einstein equation. Therefore we had to restrict ourself to cosmography. Thanks to the six symmetries and some physical requirements we still were able to identify new  degrees of freedom, that might improve our understanding of cosmological observations.

In this appendix we would like to explore the new degrees of freedom that we obtain when we try to describe the gravitational field of a static, spherically symmetric star by a torsionless connection. Here we only have four symmetries, but we are in vacuum where Einstein's equation (with vanishing cosmological constant) still makes sense. Again we will add some physical requirements:
\begin{itemize}\item
parity conservation, which is automatic for the Levi-Civita connection,
\item
asymptotic flatness of the vacuum solution, which is automatic for the Levi-Civita connection. In this asymptotic space we assume special relativity to hold. I.e there we do have a Lorentzian metric of signature $+---$. The coordinate $t$ singled out by the staticity requirement is supposed to be time-like with respect to this Lorentzian metric.
\item
We also want our geodesics to describe an attractive force admitting freely falling particles on circles.
\end{itemize}

Again our starting point is equation (\ref{killing}) describing the transformation of the connection under infinitesimal diffeomorphisms, i.e. vector fields. Here we look for all simultaneous solutions with $\xi =\pa_t$ and $\xi =J_i$. In polar coordinates the most general such solution, which is at the same time parity even, has the following non-vanishing components (up to symmetry in the last two indices):
\begin{align}
&{\Gamma ^t}_{tr}=D,&&{\Gamma ^r}_{tt}=E,\qq{\Gamma ^r}_{rr}=F,& &{\Gamma ^r}_{\theta \theta }= {\Gamma ^r}_{\varphi \varphi }/\sin^2\theta =Y,\\
& {\Gamma ^\theta }_{r\theta }= {\Gamma ^\varphi }_{r\varphi }=X,& 
 &{\Gamma ^\theta }_{\varphi \varphi  }=-\sin\theta \cos\theta ,&
  &{\Gamma ^\varphi  }_{\theta  \varphi  }= \cos\theta/\sin\theta &
\end{align}
with five functions of the radius $r$ only: $D,\,E,\,F,\,Y,$ and $X$. For the Levi-Civita connection of the metric $\dee\tau^2=B(r)\,\dee t^2-A(r)\,\dee r^2-r^2\dee\theta ^2-r^2\sin^2\theta \dee\varphi ^2$, we have: $D={\textstyle\frac{1}{2}} B'/B,\ E={\textstyle\frac{1}{2}} B'/A,\ F={\textstyle\frac{1}{2}} A'/A,\ Y=-r/A,\ X=1/r$. The prime denotes the derivative with respect to $r$.

If we want an attractive force we must have $E$ positive and for circles with constant angular velocity to exist $Y$ must be negative.

Up to anti-symmetry in the last two indices, the Riemann tensor has the following non-vanishing components:
\begin{align}
&{R^t}_{rtr}=-D'-\,(D-F)\,D,&
&{R^t}_{\theta t\theta }={R^t}_{\varphi  t\varphi }/\sin^2\theta=D\,Y ,&
\\
&{R^r}_{trt}=\qq E'-(D-F)\,E,&
&{R^\theta }_{ t\theta t}={R^\varphi }_{  t\varphi t}\qq\qq\qq=E\,X ,&
\\
&{R^r}_{\theta r\theta }= {R^r}_{\varphi  r\varphi  }/\sin^2\theta =Y'+(F-X)\,Y,&
&{R^\theta }_{ \varphi \theta \varphi }=\sin^2\theta \,[1+Y\,X],&
\\
&{R^\theta }_{ r\theta r}= {R^\varphi }_{  r\varphi  r}\qq\qq\ =-X'+(F-X)\,X,&
&{R^\varphi  }_{ \theta \varphi\theta  }=\qq\qq \ \,[1+Y\,X].&
\end{align}
 The Ricci tensor has components:
\begin{align}
&R_{tt}=E'-(D-F)\,E+2\, E\,X,&
&R_{\theta \theta }=Y'+(D+F)\,Y+1,&\\
&R_{rr}=-D'-(D-F)\,D-2X'+2\,( F-X)\,X,&
&R_{\varphi \varphi }=\sin^2\theta \,R_{\theta \theta }.&
\end{align}
Before solving the vacuum Einstein equation, let us simplify the connection by two appropriate coordinate transformations. To this end we need the transformation law of the connection under a finite  coordinate transformation $\bar x^{\bar \mu} (x)$ with Jacobi matrix
\be
{\Lambda ^{\bar \mu }}_\mu \dpp =\,\frac{\pa\bar x^{\bar\mu }}{\qq\pa x^\mu }\,.
\ee
It reads:
\be
{\bar \Gamma ^{\bar \lambda }}_{\bar\mu \bar\nu}=
{\Lambda ^{\bar \lambda }}_\lambda{ \lp\Lambda ^{-1\,T}\rp_{\bar\mu}}^\mu { \lp\Lambda ^{-1\,T}\rp_{\bar\nu}}^\nu \,
{\Gamma ^\lambda }_{\mu\nu}
+{\Lambda ^{\bar \lambda }}_\lambda{ \lp\Lambda ^{-1\,T}\rp_{\bar\mu}}^\mu\pa_\mu { \lp\Lambda ^{-1\,T}\rp_{\bar\nu}}^\lambda .
\ee
Consider the coordinate transformation with two functions $f(r)$ and $g(r)$,
\begin{align}
&\bar t\dpp =t+f(r),&&
\bar r\dpp = g(r),&&
\bar \theta \dpp =\theta ,&&
\bar \varphi \dpp = \varphi &,
\end{align}
and with Jacobian
\be
\begin{pmatrix}1&f'&0&0\\
0&g'&0&0\\
0&0&1&0\\
0&0&0&1
\end{pmatrix}.
\ee
In the barred coordinates, our connection has components given by the five functions:
\begin{align}
&\bar D=\,\frac{1}{g'}\, D\,-\,\frac{(f')^2}{g'}\, E,&
&\bar E=g'\,E&\\[2mm]
&\bar F=\,\frac{(f')^2}{g'}\, E\,+\,\frac{1}{g'}\,F\,-\,\frac{g''}{(g')^2}\,, &
&\bar Y=g'\,Y,\qq\qq \bar X=\,\frac{1}{g'}\,X.&
\end{align}
Note that from ${R^\varphi  }_{ \theta \varphi\theta  }=[1+Y\,X]$ and the negativity of $Y$ we conclude that $X$ is positive.  Otherwise we could not impose flatness asymptotically, as $r\rightarrow \infty$.
Then we use a first coordinate transformation with $f=0$ and an appropriate function $g$ to achieve $\bar X(\bar r)=1/\bar r$. Let us drop the bars and define the positive function $A\dpp=-r/Y$

We use a second coordinate transformation with $g(r)=r$ and an appropriate function $f$. Then we obtain
\begin{align}
&\bar A=A,& &\bar E=E,& &\bar D=D-(f')^2E,& &\bar F=F+(f')^2E,&
\end{align}
which motivates the definition $C\dpp=D+F$ with $\bar C=C$. We choose $f$ such that 
\be
\bar D- \bar F=\,\frac{r_S}{\bar r\,(\bar r-r_S)}\, ,
\ee
with a positive constant $r_S$, the ``Schwarzschild radius''. Again we drop the bars. 

Thanks to the two coordinate transformations we remain with only three functions, $A,\,E,$ and $C$, the first two being positive. For $D$ and $F$ we have:
\begin{align}
&D={\textstyle\frac{1}{2}} \lp C\,+\,\frac{r_S}{ r\,( r-r_S)}\,\rp,&&
F={\textstyle\frac{1}{2}} \lp C\,-\,\frac{r_S}{ r\,( r-r_S)}\,\rp.&
\end{align}
Now we integrate the three Einstein equations in vacuum:
\be
R_{tt}=E'-\lp\frac{r_S}{ r\,( r-r_S)}\,-\,\frac{2}{r}\, \rp E=0
\ee
yields
\be
E={\textstyle\frac{1}{2}} \,k\,r_S\,\frac{r-r_S}{r^3}\,,
\ee
with a dimensionless integration constant $k$. Likewise
\be
R_{rr}=-{\textstyle\frac{1}{2}} \,C'-{\textstyle\frac{1}{2}} \,\frac{-2r+3r_S}{r\,(r-r_S)}\,C=0
\ee
yields
\be
C=\,\frac{\lambda}{r_S^3}\,\frac{r^3}{r-r_S}\, ,
\ee
with another dimensionless integration constant $\lambda$.
Substituting this solution we find
\be
{R^t}_{rtr}=\,\frac{r_S}{r\,(r-r_S)}\, -{\textstyle\frac{1}{2}} \,\frac{\lambda}{r_S ^3}\,\frac{r^2}{r-r_S}\,  ,
\ee
which tends to zero for large $r$ only if $\lambda=0$. Therefore $C$ vanishes identically. Finally
\be
R_{\theta \theta }=Y'+1=0,
\ee
yields $Y=-r+(1-\ell )r_S$ with another dimensionless integration constant $\ell$ and 
\be
A=\,\frac{1}{1-(1-\ell)\,r_S/r}\, .
\ee
We still have to verify that, for this solution, all components of the Riemann tensor vanish in the limit of large $r$, which is true.

We recover the Levi-Civita connection of the (exterior) Schwarzschild solution by setting $k=1$ and $\ell=0$. 

\subsection*{Classical tests}

Of course we want to know how these two parameters are constrained by the classical tests of general relativity. To this end we have to integrate the geodesic equations. Without loss of generality, we work in the plane $\theta =\pi /2$ and using the abbreviation $B(r)\dpp = 1-r_S/r$ and the over-dot for the derivative with respect to the affine parameter $p$, we remain with:
\begin{align}
&\ddot t + \,\frac{B'}{B}\, \dot t\,\dot r\,=0,&\\[2mm]
&\ddot r+ \,\frac{k}{2}\, B'\,B\,\dot t ^2-{\textstyle\frac{1}{2}} 
\,\frac{B'}{B}\, \dot r^2\,-\, r\lp1-(1-\ell)\,\frac{r_S}{r}\rp=0,&\\[2mm]
&\ddot \varphi +\,\frac{2}{r}\, \dot r\,\dot \varphi \,=0.&
\end{align}
As in the metric case these three equations integrate once immediately:
\begin{align}
&\dot t=1/B,\qq\qq\dot \varphi =J/r^2, &\\[2mm]
&\,\frac{1}{B}\, \dot r^2\,-\,\frac{1}{B}\, +\,J^2 \gamma=\,-\epsilon\,-(k-1)\,+\,\frac{k-1}{B}\, , \label{again}
\end{align} 
with two integration constants $J$ and $\epsilon$ and the shorthand,
\begin{align}
\gamma (r)\dpp=
\,\frac{1-\ell}{r^2}\, -\,2\,\frac{\ell}{r_S^2} \lb\ln\!\lp1-\,\frac{r_S}{r}\rp+\,\frac{r_S}{r}\rb\, .
\end{align}
 The third integration constant has been set to one by a proper choice of the units of time. As declared at the beginning of the appendix, we rescale the time variable in such a way that, asymptotically, $\frac{dr}{dt}\rightarrow 1$ for a radial light ray. From the above equation, this means that photons have $\epsilon=0$, and massive particles have positive $\epsilon$.
We will have to compute the shape, $\varphi (r)$, of the trajectory. With $\dee r/\dee \varphi =\dot r/\dot \varphi $  we obtain
\begin{align}
\lp\frac{\dee r}{\dee\varphi }\rp^2\,\frac{J^2}{r^4B}\,= -\epsilon -(k-1) +\,\frac{k}{B}\, -J^2\gamma . \label{star}
\end{align}

\subsubsection*{Bending of light}

Consider now the trajectory of photons, equation ({\ref{star}) with vanishing $\epsilon$.
Let us trade the integration constant $J$ for the perihelion $r_0$ characterised by $\dee r/\dee \varphi =0$. Using equation (\ref{star}) and $\epsilon=0$ we have 
\begin{align}
J^2=\lp1-k+\,\frac{k}{B_0}\rp/\gamma _0. \label{peri}
\end{align}
We have set $B_0\dpp=B(r_0)$ and $\gamma _0\dpp=\gamma (r_0)$. We replace $J$ in equation (\ref{star}), solve for $\dee \varphi /\dee r$ and integrate:
\begin{align}
\varphi =\pm\int \,\frac{\dee r}{r^2\sqrt{B}}\, 
\lb
\frac{k/B+1-k}{k/B_0+1-k}\,\gamma _0\,-\,\gamma \,\rb^{-1/2}.
\end{align}
To do the integral, we linearize the integrand in $r_S/r_0$ and set $x\dpp=r/r_0$. Using 
\begin{align}
\gamma \sim\,\frac{1}{r^2}\, \lb1+\,\frac{2\ell}{3}\,\frac{r_S}{r}\,\rb,  
\end{align}
we find:
\begin{align}
\varphi \sim\pm\int \,\frac{\dee x}{x}\,\frac{1}{\sqrt{x^2-1}}\,  
\lb 1+{\textstyle\frac{1}{2}} \,\frac{r_S}{r_0}\,\frac{1}{x}\,+
{\textstyle\frac{1}{2}}  k\,\frac{r_S}{r_0}\,\frac{x}{x+1}\, -
{\textstyle\frac{1}{3}}  \ell\,\frac{r_S}{r_0}\,\frac{x^2+x+1}{x\,(x+1)}\, \rb=\dpp\int I(x)\,\dee x.\label{phi}
\end{align}
  Let us choose the incoming direction of the photons at $\varphi =\pi $, denote the scattering angle by $\Delta \varphi $ and integrate 
equation (\ref{phi}) on the left-hand side from $\varphi= \varphi _0$ to $\pi $ and on the right-hand side from $x=1$ to $\infty$. In this domain $\dee\varphi /\dee r$ is positive and we have:
\begin{align} \Delta \varphi =\pi -2\varphi _0=\pi -2\lp\pi -\int_1^\infty I(x)\,\dee x\rp=\,\frac{r_S}{r_0}\,\lb1+k-\,\frac{4}{3}\, \ell\rb. 
\end{align}
We have used the following integrals:
\begin{align}
\int\,\frac{\dee x}{x\sqrt{x^2-1}}\,&=-\arcsin\,\frac{1}{x}\,,&
 \int\,\frac{\dee x}{x^2\sqrt{x^2-1}}\,&=\,\frac{\sqrt{x^2-1}}{x}\,,&\\[2mm]
  \int\,\frac{\dee x}{(x+1)\sqrt{x^2-1}}\,&=\,\sqrt{\frac{x-1}{x+1}}\,,&
  \int\,\frac{\dee x}{\sqrt{x^2-1}}\,\frac{x^2+x+1}{x^2(x+1)}\, &=\,\sqrt{x^2-1}\lp\,\frac{1}{x}\,+\,\frac{1}{x+1}\,\rp  \,.&\label{int}
\end{align}
2010 data from Very Long Baseline Interferometry \cite{will} constrain the parameters at the $10^{-4}$ level:
\begin{align}
(k-1)-\,\frac{4}{3}\, \ell =(-0.8\pm1.2)\cdot10^{-4}.
\end{align} 
On galactic scales, let us mention data from lensing by fifteen elliptical galaxies in the Sloan Digital Sky Survey \cite{bolt}. They constrain our parameters at the 10 \%  level.

\subsubsection*{Time delay of light}

Our starting point is again equation (\ref{again}) with $\epsilon=1-k$ and $J$ expressed in terms of the perihelion $r_0$, equation (\ref{peri}). Here we need $t(r)$ noting that $t$ is the proper time of an observer at rest at infinity:
\begin{align}
\,\frac{\dee t}{\dee r}\, =\,\frac{\dot t}{\dot r}\, =
\,\frac{1}{B}\, \lb (k+(1-k)\,B)\,- (k+(1-k)\,B_0)\,\frac{B\gamma }{B_0\gamma _0}\,\rb^{-1/2}.
\end{align}
We linearize in $r_S/r_0$,
\bb
\,\frac{\dee t}{\dee r}\,\sim\,\frac{x}{\sqrt{x^2-1}}\, \lb1+\,\frac{1}{x\,(x+1)}\, \la x+\,\frac{3}{2}\,  -\,\frac{k-1}{2}\,x-\,\frac{\ell}{3}\,\ra 
\,\frac{r_S}{r_0}\,\rb.
\eee
Integrating we obtain the time of flight of the photon from a location $r$ to the perihelion,
\begin{align}
t(r,r_0)\sim \sqrt{r^2-r_0^2}+r_S\lb\lp \frac{1}{2}\, -\,\frac{k-1}{2}\,-\,\frac{\ell}{3}\rp\frac{\sqrt{x-1}}{\sqrt{x+1}}\, 
+\lp1 -\,\frac{k-1}{2}\rp\ln(x+\sqrt{x^2-1}\,)\rb,
\end{align}
where we have used the first integral in (\ref{int}) and
\begin{align}
\int\,\frac{x\,\dee x}{(x+1)\sqrt{x^2-1}}\,
=-\,\frac{\sqrt{x-1}}{\sqrt{x+1}}\, +\ln(x+\sqrt{x^2-1}\,).
\end{align}
The time delay after a round trip of the photon from the earth at $r_e$ to the reflector at $r_r$ and back therefore is:
\begin{align}
\Delta t\sim
2\,r_S\lb\lp \frac{1}{2}\, -\,\frac{k-1}{2}\,-\,\frac{\ell}{3}\rp
\la
\frac{\sqrt{r_e^2-r_0^2}}{\sqrt{r_e^2+r_0^2}}\, 
+\,\frac{\sqrt{r_r^2-r_0^2}}{\sqrt{r_r^2+r_0^2}}
\right.\ra
\qq\qq\qq\qq\qq\qq\qq
\nonumber\\[2mm]\left.
+\lp1 -\,\frac{k-1}{2}\rp\la \ln\,\frac{r_e+\sqrt{r_e^2-r_0^2}}{r_0}\, 
+\ln\,\frac{r_r+\sqrt{r_r^2-r_0^2}}{r_0}\ra 
\rb.
\end{align}
The most precise measurement of the time delay today comes from the Doppler tracking of the Cassini spacecraft  in 2002 \cite{bert} \cite{will}, with $r_0=1.6\,R_\odot$ and $r_r=6.43$ AU. The precision of this measurement is at the $10^{-5}$ level and yields for our parameters:
\begin{align}
11.0\,(k-1)-1.3\, \ell =(-2.1\pm2.3)\cdot10^{-5}.
\end{align}

\subsubsection*{Perihelion advance of Mercury}

Let us denote by $r_+$ and $r_-=r_0$ the maximum and minimum distance between Mercury and sun. The orbit of Mercury is given by
 equation ({\ref{star}) with positive $\epsilon$.
Let us trade the two integration constants $J$ and $\epsilon$ for  $r_+$ and $r_-$ characterised by $\dee r/\dee \varphi =0$. Equation (\ref{star}) yields:
\begin{align}
J^2=k\,\frac{1/B_+-1/B_-}{\gamma_+-\gamma _-}\,\qq {\rm and}\qq
\epsilon=k\,\frac{\gamma _+/B_+-\gamma _+/B_-}{\gamma_+-\gamma _-}\,-(k-1),
\end{align}
Replacing $J$  and $\epsilon$ in equation (\ref{star}) we see that the $k$s cancel. Solving for $\dee\varphi /\dee r$ and integrating we have:
\begin{align}
\varphi (r_+)=\dpp \varphi _+=\varphi _-+\int_{r_-}^{r_+}\,\frac{\dee r}{r^2\sqrt{B}}\, \lb
\,\frac{(1-B/B_-)\gamma _+-(1-B/B_+)\gamma _-}{B/B_+-B/B_-}\, 
-\gamma \,\rb.
\end{align}
As usual we linerize the integrand in $r_S/r_-$,
\begin{align}
\varphi _+-\varphi _-=&
\lb \sqrt{r_+r_-}\,+\,\frac{3-2\ell}{6}\,\frac{r_++r_-}{\sqrt{r_+r_-}}\, r_S\,\rb
\int_{r_-}^{r_+}\frac{\dee r}{r\sqrt{(r_+-r)(r-r_-)}}\nonumber\\[2mm]
&\qq\qq\qq+\,\frac{3-2\ell}{6}\,{\sqrt{r_+r_-}}\, r_S\,
\int_{r_-}^{r_+}\frac{\dee r}{r^2\sqrt{(r_+-r)(r-r_-)}}.
\end{align}
Using the integrals,
\begin{align}
\int\frac{\dee r}{r\sqrt{(r_+-r)(r-r_-)}}=&\,\frac{2}{\sqrt{r_+r_-}}\, \arctan\sqrt{\,\frac{r_+}{r_-}\,\frac{r-r_-}{r_+-r}}\,, \\[2mm]
 \int\frac{\dee r}{r^2\sqrt{(r_+-r)(r-r_-)}}=&\,\frac{1}{r_+r_-}\, 
\frac{\sqrt{(r_+-r)(r-r_-)}}{r}\, \nonumber\\&+
 \,\frac{r_++r_-}{\sqrt{r_+r_-}^3}\, \arctan\sqrt{\,\frac{r_+}{r_-}\,\frac{r-r_-}{r_+-r}}\,,
 \end{align}
we obtain
\begin{align}
\varphi _+-\varphi _-=\pi \lb\,1+\,\frac{3-2\ell}{4}\,\frac{r_++r_-}{r_+r_-}\,r_S\,\rb\,.
\end{align}
Finally the perihelion shift $\Delta \varphi $ after one circumnavigation is given by \cite{weinberg}:
\begin{align}
\Delta \varphi=2(\varphi _+-\varphi _-)-2\pi =\,\frac{3-2\ell}{2}\,\frac{r_++r_-}{r_+r_-}\,r_S.\label{comp}
\end{align}
Today, the most precise measurement of the perihelion advance include 2013 data on the orbit of Mercury collected by the spacecraft Messenger that orbited Mercury \cite{ver}\cite{will}. Comparison with equation (\ref{comp}) yields,
\begin{align}
\ell=(-0.15\pm 1.3)\cdot 10^{-5}.
\end{align}

\subsubsection*{Gravitational redshift}

For this test we would need a proper time not only far away from the sun, but everywhere in space. We do not see a natural substitute  for the metric, that would define this proper time using the connection only. However within $10^{-5}$, the other three classical tests constrain our connection to derive from a metric. Using this metric to define proper time is of course what Einstein did a century ago.

\end{document}